\documentclass[runningheads]{llncs}
\usepackage[T1]{fontenc}
\usepackage{graphicx}
\usepackage{amsmath}
\usepackage{algorithm}
\usepackage{algpseudocode}
\algnewcommand{\LeftComment}[1]{\(\triangleright\) #1}

\usepackage{xcolor}
\usepackage{booktabs}  
\usepackage{multirow} 
\usepackage{hyperref}
\begin{document}
\title{SARHAchat: An LLM-Based Chatbot for Sexual and Reproductive Health Counseling}
\titlerunning{SARHAchat: Sexual
And Reproductive Health Assistant Chatbot}
%
\author{Jiaye Yang\orcidID{0009-0008-7220-587\textsc{X}} \and
Xinyu Zhao\orcidID{0009-0000-0253-5488} \and \\
Tianlong Chen\textsuperscript{*}\orcidID{0000-0001-7774-8197} \and \\
Kandyce Brennan\textsuperscript{*}\orcidID{0000-0001-8729-7828}}
\authorrunning{J. Yang et al.}
%
\institute{
University of North Carolina at Chapel Hill, Chapel Hill NC 27514, USA\\
\email{yamseyoung@gmail.com} \space \email{\{xinyu,tianlong\}@cs.unc.edu} \space \email{heddrick@unc.edu}}

\renewcommand{\thefootnote}{\fnsymbol{footnote}} 
\footnotetext[1]{Equal corresponding author.}

\maketitle              

\begin{abstract}
While Artificial Intelligence (AI) shows promise in healthcare applications, existing conversational systems often falter in complex and sensitive medical domains such as Sexual and Reproductive Health (SRH). These systems frequently struggle with hallucination and lack the specialized knowledge required, particularly for sensitive SRH topics. Furthermore, current AI approaches in healthcare tend to prioritize diagnostic capabilities over comprehensive patient care and education. Addressing these gaps, this work at the UNC School of Nursing introduces SARHAchat, a proof-of-concept Large Language Model (LLM)-based chatbot. SARHAchat is designed as a reliable, user-centered system integrating medical expertise with empathetic communication to enhance SRH care delivery. Our evaluation demonstrates SARHAchat's ability to provide accurate and contextually appropriate contraceptive counseling while maintaining a natural conversational flow. The demo is available at \hyperlink{https://sarhachat.com/}{https://sarhachat.com/}.

\vspace{-3mm}
\end{abstract}
\vspace{-3mm}

\section{Introduction}
\vspace{-3mm}
Artificial Intelligence (AI) is significantly reshaping healthcare delivery, enhancing efficiency and personalization across applications such as disease diagnosis, drug discovery, and mental health support \cite{gabriel2024ai,varoquaux2022machine,ye2024role,hou2024selfexplainableaimedicalimage,torous2024generative}. Conversational chatbots powered by Large Language Models (LLMs) offer considerable potential for assisting patients and providers with tasks such as medical question answering and summarizing health records \cite{singhal2025toward,vanveen2024adapted}. However, deploying these powerful tools safely and effectively in high-stakes healthcare settings necessitates robust safeguards and rigorous evaluation, particularly given two prevalent challenges. First, current healthcare chatbots often struggle with \textbf{hallucination}, the tendency to generate inaccurate or fabricated information. Second, many systems lack sufficient focus on crucial \textbf{user-centered aspects}, failing to adequately address end-user needs related to emotional support, trust, fairness, and health literacy \cite{laymouna2024roles}. While there's a trend towards using chatbots for initial symptom checking and diagnostics \cite{tu2024towards}, less attention has been paid to non-diagnostic conversational roles, such as discussing patient preferences or engaging in open-ended counseling to determine appropriate recommendations.

Motivated by these observations,
this work introduces SARHAchat (\textbf{S}exual \textbf{A}nd \textbf{R}eproductive \textbf{H}ealth \textbf{A}ssistant \textbf{chat}bot), an initiative from the UNC School of Nursing. SARHAchat aims to provide reliable AI-driven patient triaging and counseling for SRH topics. It facilitates pre-clinical contraceptive care by gathering user health histories, delivering personalized counseling and education, and creating user profiles to streamline subsequent provider interactions and prescription processes. Key contributions include:

$\bullet$ We equip the chatbot to recommend suitable contraceptive methods following predefined 5-stage flow, and incorporating Retrieval-Augmented Generation (RAG) and prompting techniques to reduce hallucinations and ensure contextually appropriate, diverse responses.

$\bullet$ We implement a generative user interface that supports seamless, multimodal interaction, combining textual responses with visual aids and downloadable resources to enhance comprehension and information delivery.

\begin{figure}[t]
\includegraphics[width=\textwidth]{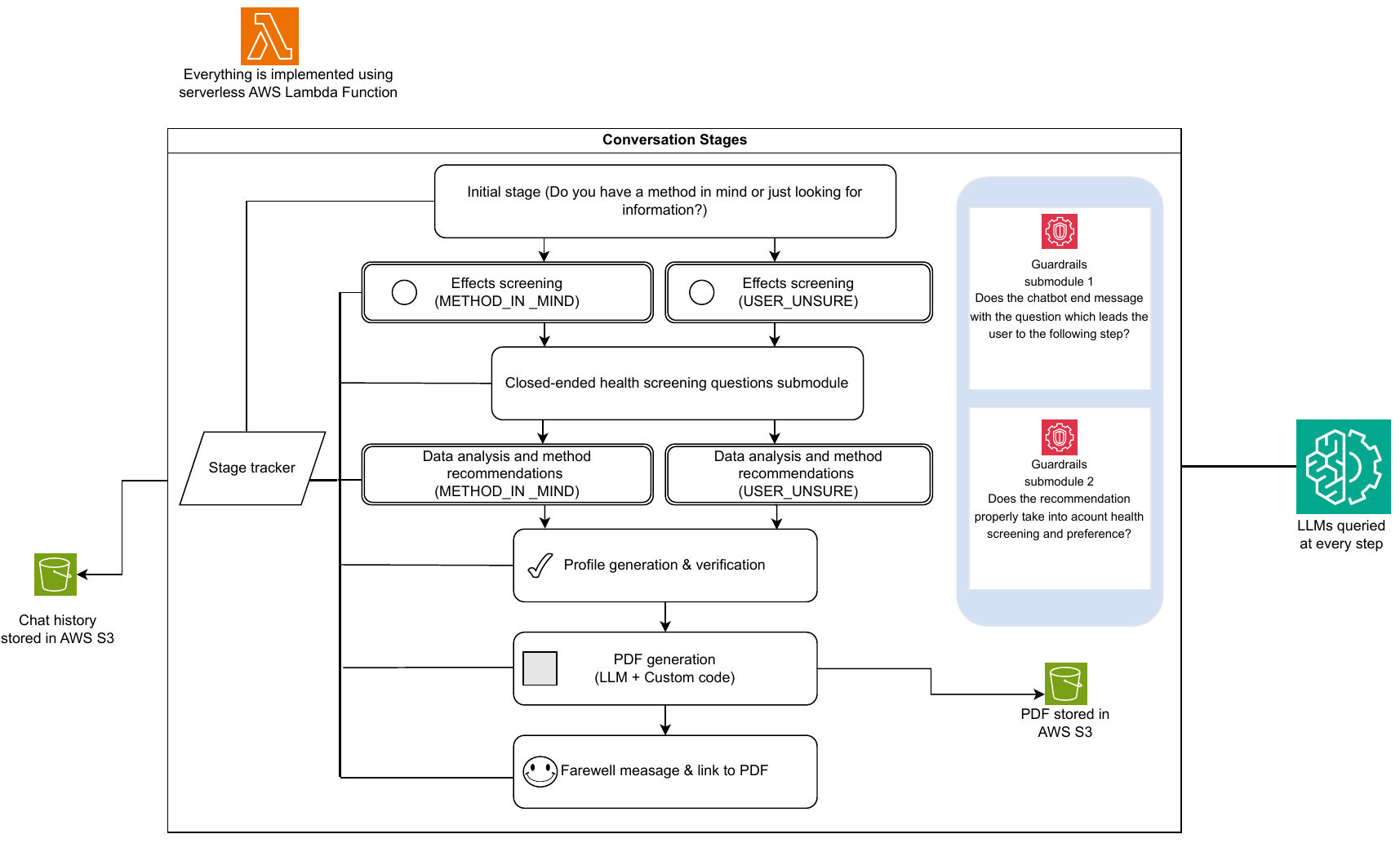}
\vspace{-6mm}
\caption{Overview of the SARHAchat conversation logic and system components. } \label{fig1}
\vspace{-6mm}
\end{figure}

\renewcommand{\thefootnote}{\arabic{footnote}}

\vspace{-5mm}
\section{Overview of SARHAchat}
\label{sec:system_arch}
\vspace{-4mm}
The SARHAchat framework, depicted in Figure~\ref{fig1}, builds upon LLM-based chatbot architecture \cite{xi2023risepotentiallargelanguage} integrating a memory module and a guided recommendation algorithm. Specifically, we adopt GPT-4\footnote{https://platform.openai.com/docs/models/gpt-4-turbo} as the LLM backbone. The system guides users through a predefined five-stage conversational flow: (1) initial information gathering (intent, gender, experience), (2) preference screening (frequency, side effects), (3) health screening (medical history), (4) appropriate contraception recommendation, and (5) profile verification and generation of a downloadable summary. Stage-specific prompts initiate questions within each phase, while a stage tracker monitors the conversation state and manages transitions based on user responses.

\vspace{-3mm}
\section{Methodology}
\vspace{-3mm}

\paragraph{\textbf{Bi‑level Memory Module.}}
To manage conversational context and ensure factual accuracy beyond the LLM's inherent limitations, SARHAchat utilizes a Bi‑level memory system. \textbf{Short-term memory} maintains the immediate dialogue history, dynamically extracting user factors to query relevant information from a structured long-term knowledge. Meanwhile, \textbf{external long-term memory} is organized as an expert-maintainable key-value store containing distilled, up-to-date knowledge on contraceptive methods (\textit{e.g.}, incorporating current CDC guidelines such as the 2024 MEC\footnote{U.S. Medical Eligibility Criteria for Contraceptive Use, 2024. \url{https://www.cdc.gov/mmwr/volumes/73/rr/rr7304a1.htm}} and SPR\footnote{U.S. Selected Practice Recommendations for Contraceptive Use, 2024 (U.S. SPR). \url{https://www.cdc.gov/mmwr/volumes/73/rr/rr7303a1.htm}}), overcomes LLM context window constraints. This architecture allows the integration and retrieval of verified, current information throughout the conversation, enhancing reliability and reducing reliance solely on the LLM's pre-training data.

\vspace{-3mm}
\paragraph{\textbf{Structured Reasoning and Guided Response Generation.}}
Collaboratively designed with healthcare professionals, our chatbot employs a step-by-step reasoning chain that mimics expert clinical decision-making, explicitly considering user preferences, medical history, and weighted decision criteria relevant to contraceptive choice. This structured process simplifies the reasoning task, ensuring the LLM's response remains controllable and grounded within the scope of verified information. Subsequently, a "thought injection" prompting technique integrates this reasoning chain and an explicit rationale into the final prompt, guiding the LLM to generate accurate, contextually appropriate, and justified recommendations in natural language, thereby mitigating hallucination risk while maintaining conversational fluidity over hard-coded responses.

\vspace{-4mm}
\section{Evaluation}
\vspace{-3mm}
Evaluation data comprised two types: complete chat conversation histories and user feedback questionnaires collected via Qualtrics post-interaction. The conversation dataset is gathered from two sources: (1) interactions generated during controlled beta testing involving nursing and healthcare professionals, and (2) synthetic dialogues simulating diverse fictional patient actors (varied demographics, medical histories, lifestyles) across a breadth of scenarios. Medical experts annotated these chat histories, assessing both information quality and bedside manner. The complete dialogue corpus, thoroughly redacted to remove personally identifiable information (PII), is maintained in a proprietary cloud database.

The evaluation dfocuses on both the clinical capability validation and subjective user experience covering two essential dimensions: medical safety and conversational quality of the chatbot. 
\vspace{-3mm}
\paragraph{\textbf{Medical safety.}} This metric assesses the accuracy and appropriateness of the chatbot's recommendations. Performance is measured via pass/fail assessments for critical functions. Qualitative evaluation categorizes failure cases into three main types: (1) omission of essential health screening questions; (2) recommendation of contraindicated contraceptive methods; and (3) errors or incompleteness in critical information provided about contraceptive methods, such as side effects or contraindications.

\vspace{-3mm}
\paragraph{\textbf{Conversational quality.}} This metric evaluates the chatbot's ability to engage in natural, coherent, and contextually appropriate dialogue simulating a clinical interview. We assess conversations using rubric-based graded scores based on: (1) Interaction Coherence, encompassing logical flow, clarity, appropriate summarization, and avoidance of redundancy; and (2) Empathetic Engagement, reflecting appropriate tone and demonstrated understanding and support for user concerns. A conversation receives a ``Satisfactory'' rating only if performance across both dimensions met the predefined standards outlined in the rubric.
\vspace{-2mm}

\begin{table}[htbp]
\centering
\vspace{-2mm}
\caption{System Performance Comparison (N=169 conversations per version)}
\vspace{-2mm}
\begin{tabular}{@{}lcc@{}}
\toprule
Metric Category & Baseline & SARHAchat \\
\midrule
\multicolumn{3}{l}{\textbf{Medical Safety}} \\
\quad Pass Rate                 & 85.21\% & 98.22\% \\
\multicolumn{3}{l}{\quad \textit{Failure Analysis}} \\
\quad - Health Screening Omission  & 1 & 0 \\
\quad - Contraindicated Method & 13 & 0 \\
\quad - Critical Information Error/Incompleteness & 11 & 3 \\
[0.5em]
\multicolumn{3}{l}{\textbf{Conversational Quality}} \\
 \quad Satisfactory             & 151 (89.35\%) & 167 (98.82\%) \\
 \quad Needs Improvement           & 18 (10.65\%) & 2 (1.18\%) \\
\bottomrule
\multicolumn{3}{l}{\small Note: Results based on expert evaluation of dialogue samples.}
\end{tabular}
\label{tab:performance}
\end{table}
\vspace{-5mm}

We evaluate the proposed SARHAchat against a previous version (Baseline), which is implemented by naive prompting ChatGPT with user queries and documents. As shown in Table 1, SARHAchat demonstrate superior Medical Safety, achieving a notably higher pass rate (98.22\% vs 85.21\%) and drastically reducing critical failures. Specifically, health screening omissions and contraindicated method suggestions are barely observed, and critical information errors decrease from 11 to 3. Regarding Conversational Quality, SARHAchat also performed better, receiving significantly more satisfactory ratings (98.82\% vs 89.35\%) and substantially less feedback requiring improvement.


\vspace{-4mm}
\section{Conclusion}
\vspace{-3mm}
This paper introduced SARHAchat, an LLM-based chatbot that pioneers in applying AI to Sexual and Reproductive Health.
SARHAchat facilitates reliable patient counseling through a intuitive chat interface, underpinned by a framework integrating comprehensive medical knowledge via an bi-level memory module with patient-centered communication strategies. 
Our methodology employs structured reasoning and guided response generation to deliver personalized, accurate recommendations grounded in evidence-based guidelines, effectively mitigating risks such as hallucination and ensuring responses remain aligned with verified information.
Evaluation indicates SARHAchat streamlines SRH pre-clinical care and demonstrates strong potential for clinical deployment with minimal technical barriers. 


%
%
%
\bibliographystyle{splncs04}
\bibliography{mybibliography}






\end{document}